\documentclass[10pt,letterpaper]{article}
\usepackage[top=0.85in,left=2.75in,footskip=0.75in]{geometry}

\usepackage{amsmath,amssymb}

\usepackage{changepage}

\usepackage{textcomp,marvosym}

\usepackage{cite}

\usepackage{nameref,hyperref}

\usepackage[nopatch=eqnum]{microtype}
\DisableLigatures[f]{encoding = *, family = * }

\usepackage[table]{xcolor}

\usepackage{array}

\newcolumntype{+}{!{\vrule width 2pt}}

\newlength\savedwidth

\raggedright
\usepackage[aboveskip=1pt,labelfont=bf,labelsep=period,justification=raggedright,singlelinecheck=off]{caption}

\usepackage{graphicx}%
\usepackage{multirow}%
\usepackage{amsmath,amssymb,amsfonts}%
\usepackage{amsthm}%
\usepackage{mathrsfs}%
\usepackage[title]{appendix}%
\usepackage{xcolor}%
\usepackage{textcomp}%
\usepackage{manyfoot}%
\usepackage{booktabs}%
\usepackage{algorithm}%
\usepackage{algorithmicx}%
\usepackage{algpseudocode}%
\usepackage{listings}%
\usepackage{paralist, tabularx}%

\makeatletter
\renewcommand{\@biblabel}[1]{\quad#1.}
\makeatother

\usepackage{lastpage,fancyhdr,graphicx}
\usepackage{epstopdf}
\fancyheadoffset[L]{2.25in}
\fancyfootoffset[L]{2.25in}
\begin{document}
\vspace*{0.2in}

\begin{flushleft}
{\Large
\textbf\newline{Can OpenAI o1 outperform humans in higher-order cognitive thinking?} 
}
\newline
\\

Ehsan Latif\textsuperscript{1,2},
Yifan Zhou\textsuperscript{1,2,3\Yinyang},
Shuchen Guo\textsuperscript{1,2,4\Yinyang},
Yizhu Gao\textsuperscript{1,2,5\Yinyang},
Lehong Shi\textsuperscript{2,6\Yinyang},
Matthew Nyaaba\textsuperscript{1,5\Yinyang},
Arne Bewerdorff\textsuperscript{7\Yinyang},
Xiantong Yang\textsuperscript{8\Yinyang},
Xiaoming Zhai\textsuperscript{1,2,5*}
\\
\bigskip
\textbf{1} AI4STEM Education Center, University of Georgia, Athens, Georgia, USA
\\
\textbf{2} National GENIUS Center, University of Georgia, Athens, Georgia, USA
\\
\textbf{3} School of Computing, University of Georgia, Athens, Georgia, USA
\\
\textbf{4} School of Teacher Education, Nanjing Normal University, Nanjing, Jiangsu, China
\\
\textbf{5} Department of Mathematics, Science, and Social Studies Education, University of Georgia, Athens, Georgia, USA
\\
\textbf{6} Department of Workforce Education and Instructional Technology, University of Georgia, Athens, Georgia, USA
\\
\textbf{7} School of Social Sciences and Technology, Technical University of Munich, Munich, Bavaria, Germany
\\
\textbf{8} Faculty of Psychology, Beijing Normal University, Beijing, China
\\
\bigskip

\Yinyang These authors contributed equally to this work.

*Corresponding aurhor. Email: xiaoming.zhai@uga.edu

\end{flushleft}
\section*{Abstract}
This study evaluates the performance of OpenAI’s o1-preview model in higher-order cognitive domains, including critical thinking, systematic thinking, computational thinking, data literacy, creative thinking, logical reasoning, and scientific reasoning. Using established benchmarks, we compared the o1-preview models’s performance to human participants from diverse educational levels.  o1-preview achieved a mean score of 24.33 on the Ennis-Weir Critical Thinking Essay Test (EWCTET), surpassing undergraduate (13.8) and postgraduate (18.39) participants (\textit{z} = 1.60 and 0.90, respectively). In systematic thinking, it scored 46.1 ± 4.12 on the Lake Urmia Vignette, significantly outperforming the human mean (20.08 ± 8.13, \textit{z} = 3.20). For data literacy, o1-preview scored 8.60 ± 0.70 on Merk et al.’s “Use Data” dimension, compared to the human post-test mean of 4.17 ± 2.02 (\textit{z} = 2.19). On creative thinking tasks, the model achieved originality scores of 2.98 ± 0.73, higher than the human mean of 1.74 (\textit{z} = 0.71). In logical reasoning (LogiQA), it outperformed humans with 90\% ± 10 accuracy versus 86\% ± 6.5 (\textit{z} = 0.62). For scientific reasoning, it achieved near-perfect performance (0.99 ± 0.12) on the TOSLS,, exceeding the highest human scores of 0.85 ± 0.13 (\textit{z} = 1.78). While o1-preview excelled in structured tasks, it showed limitations in problem-solving and adaptive reasoning. These results demonstrate the potential of AI to complement education in structured assessments but highlight the need for ethical oversight and refinement for broader applications.



\section{Introduction}

Artificial Intelligence (AI) has made significant advances in recent years, particularly with the emergence of large language models (LLMs) like OpenAI's o1-preview model \cite{openai2024o1}. These developments have created new opportunities for integrating AI into education, especially to enhance higher-order thinking skills that are essential for academic and professional success \cite{zhai2024can, guo2024artificial}. Higher-order thinking skills—such as critical thinking, systematic thinking, metacognition, logical reasoning, and collaborative problem-solving—are crucial for navigating the complexities of modern education and for developing cognitive abilities required in the 21st-century workforce \cite{lewis1993defining, collins2014skills}.

Research has underscored AI's potential to improve problem-solving and cognitive skills, including critical thinking and data literacy, with significant implications for education and research \cite{zhai2022chatgpt, zhong2024evaluationopenaio1opportunities}. Reviews and bibliometric analyses highlight a decade of progress in AI applications for education \cite{guo2024artificial}. However, critical questions remain about the extent to which AI models can replicate or surpass human performance in higher-order thinking tasks, particularly at the graduate education level \cite{zhai2024can, marino2024fast}.

Evaluations of the OpenAI o1-preview model reveal promise in addressing cognitive challenges, such as complex problem-solving and logical reasoning \cite{hu2024can, lightman2023verify}. Despite this, concerns persist about the model's ability to consistently perform tasks requiring deeper cognitive processes, such as metacognition \cite{renze2024self} and scientific reasoning \cite{zhai2024can}. Mixed results from studies on analogical reasoning in AI indicate some successes but also highlight challenges in achieving the nuanced thinking characteristic of human cognition \cite{qin2024relevant, musker2024semantic}.

\begin{figure}[htp!]
    \centering
    \includegraphics[width=0.75\linewidth]{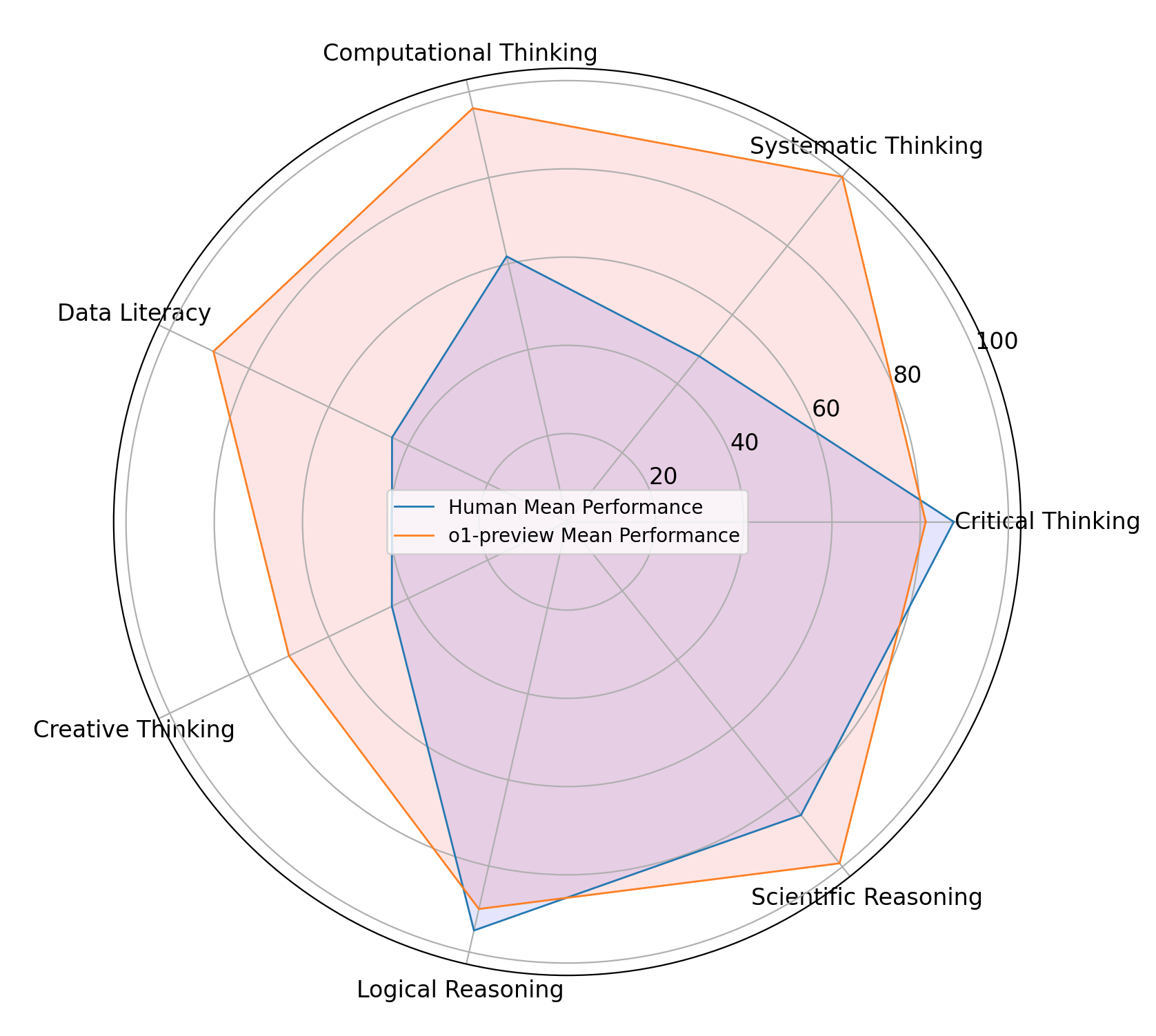}
    \caption{Performance overview of OpenAI o1-preview in higher-order thinking domains compared to human experts.}
    \label{fig:o1-preview_performance_intro}
\end{figure}

This paper addresses these gaps by comprehensively evaluating the o1-preview model across several key higher-order thinking domains, specifically \textit{critical thinking}, \textit{systematic thinking}, \textit{computational thinking}, \textit{data literacy}, \textit{creative thinking}, \textit{logical reasoning}, and \textit{scientific reasoning}. These domains are identified as critical for graduate education and beyond \cite{zhai2024can, guo2024artificial, lewis1993defining, collins2014skills}. Our findings reveal that the o1-preview model outperforms human experts in five out of seven domains, including systematic thinking, computational thinking, data literacy, creative thinking, and scientific reasoning (see Fig.~\ref{fig:o1-preview_performance_intro} for an overview of its performance).

While the o1-preview model demonstrates effectiveness in computational tasks \cite{zhai2024can, guo2024artificial}, further research is needed to evaluate its capabilities in collaborative thinking and human-AI interactions in dynamic educational contexts \cite{zhang2023exploring, lykov2024industry}. Recent studies suggest that incorporating reflective and deconstructive strategies into AI systems can enhance their performance on complex, multi-step problems \cite{lingo2024enhancing, renze2024self}. Additionally, the ethical implications of using AI in education, particularly in decision-making and reasoning tasks, demand careful consideration \cite{sorin2024ethical}.

In the following sections, we systematically evaluate the o1-preview model's performance across these higher-order thinking domains, benchmarking its results against human participants. Our analysis provides insights into the potential and limitations of AI in graduate education, identifying areas where AI can contribute to the future of learning and teaching.





\section{Methods}
\subsection{Participants}
This study utilizes OpenAI’s o1-preview model to evaluate its capacity for performing higher-order thinking tasks. Human participant data are drawn from existing studies representing various educational levels, including undergraduate, graduate, and post-graduate students. Graduate-level human performance serves as the benchmark where applicable. The performance of the o1-preview model is compared to human participants across multiple higher-order thinking domains.

\subsection{Instruments and Datasets}
This study employs well-established instruments and datasets tailored to each higher-order thinking category. For each domain, the performance of the o1-preview model is evaluated against the corresponding human cohort.

\subsubsection{Critical Thinking Instruments}
The Ennis-Weir Critical Thinking Essay Test (EWCTET) \cite{ennis1985cornell} is a widely recognized tool designed to assess critical thinking skills through performance-based, context-driven written responses. Unlike standardized assessments, the EWCTET emphasizes real-world application, making it particularly effective for evaluating higher-order cognitive abilities. Studies such as those by Taghinezhad et al. \cite{taghinezhad2018critical} highlight its capacity to assess cognitive complexity effectively. Developed by Robert Ennis and Eric Weir, the EWCTET challenges participants to construct, analyze, and evaluate arguments through coherent essays \cite{werner1991ennisweir}.

The EWCTET uses scenario-based prompts, requiring participants to critically assess specific situations or arguments and present well-reasoned evaluations in written form \cite{prayogi2018cibl}. It measures a range of critical thinking skills, including logical consistency, identification of assumptions, argument analysis, and evidence-based reasoning \cite{seker2008questioning}. The tool's accessibility and cost-effectiveness further contribute to its frequent use in educational research and practice.

The assessment typically takes 40 to 60 minutes, allowing participants sufficient time to construct thoughtful and nuanced arguments \cite{ennis1985cornell}. The scoring rubric evaluates multiple dimensions of critical thinking and written communication, including clarity, logical structure, and engagement with opposing viewpoints \cite{seker2008questioning}.

Specific scoring criteria include:

\textbf{Recognition of misuse of analogy, irrelevance, and circular reasoning:} Assesses the ability to identify logical fallacies.

\textbf{Recognition of inadequate sampling and other logical fallacies:} Evaluates awareness of reasoning flaws, such as biased evidence.

\textbf{Scoring for individual criteria:} Scores range from -1 to 3 points (poor to well-supported reasoning):

\textbf{Ninth criterion:} Evaluates the overall quality of argumentation, including summarization and recognition of emotive language, with a maximum score of 5 points.

\begin{table}[ht]
    \centering
    \small
    \caption{Summary of Human Performance on the Ennis-Weir Critical Thinking Essay Test (EWCTET)}
    \begin{tabularx}{\textwidth}{X X X X}
        \toprule
        \textbf{Authors} & \textbf{Participants} & \textbf{Assessment Type} & \textbf{Mean Score (SD)} \\ \midrule
        Hatcher \cite{hatcher2006stand} & American freshmen at Baker University (four-year study) & Post-test after one-year compulsory critical thinking course & 11.8–13.8 (SD not provided); Gains: 2.8–6.0 \\ 
        Davidson and Dunham \cite{davidson1996efl} & 36 Japanese EFL college students (treatment: n=17, control: n=19) & Post-test following critical thinking seminar for treatment group & Treatment: 6.6, Control: 0.6, $p < .001$ \\ 
        Hollis et al. \cite{hollis2024icteta} & 100 online participants (63 females, 37 males), varied education levels & Post-test, no additional training & 14.31 (SD = 8.45); Postgraduate: 18.39 (SD = 6.59), Undergraduate: 11.51 (SD = 8.23) \\ 
        \bottomrule
    \end{tabularx}
    \label{tab:ewctet_summary}
\end{table}

To evaluate the critical thinking performance of OpenAI’s o1-preview model, human performance data from studies using the EWCTET were established as a benchmark. These studies provide a comprehensive basis for assessing whether the model can match or exceed human critical thinking capabilities. Tab.~\ref{tab:ewctet_summary} summarizes the participants, assessment types, and results from these studies.

\subsubsection{Systematic Thinking Instruments}
Numerous instruments have been developed to assess systematic thinking (ST), leveraging various theoretical frameworks across disciplines, using diverse methods, and targeting different educational levels. Common formats include mapping, interviews, scenario-based items, open-ended items, fill-in-the-blank items, and multiple-choice questions \cite{dugan2021systems}. 

In this study, we selected two ST instruments based on the following criteria: 
\begin{enumerate}
    \item To account for the limitations of the o1-preview model in identifying and generating images, we excluded instruments requiring respondents to create mappings or interpret indispensable visual information in prompts.
    \item  Research supports the use of scenario-based, ill-structured items as the most effective method for measuring ST. These items present realistic problems followed by a series of open- or close-ended questions to elicit ST skills \cite{norris2022investigating}.
    \item Since ST is extensively studied in biology and engineering \cite{dugan2021systems, li2024developing}, we focused on theoretical frameworks and instruments developed in these fields, as they are more advanced and comprehensive.
\end{enumerate}

Based on these considerations, we selected two recently developed scenario-based, ill-structured instruments designed for higher education: the "Village of Abeesee" instrument and the "Lake Urmia Vignette" (LUV) instrument. These instruments are grounded in distinct theoretical frameworks and contextualized in engineering and socio-environmental systems.

\paragraph{Village of Abeesee Instrument}
The Village of Abeesee instrument, developed by Grohs et al. (2018) \cite{grohs2018assessing}, is based on a framework that views ST as comprising three dimensions for general interdisciplinary use. This framework emphasizes the interconnectedness of technical and social aspects of modern problems and highlights the importance of stakeholder perspectives. 

The instrument presents a scenario about the Village of Abeesee, which faces a complex issue regarding winter heating. Participants respond to six open-ended questions aligned with the framework's dimensions. A pilot study was conducted to refine the scenario and gather qualitative data. Rubrics were developed through a multi-stage process using a pool of 93 student responses. The scoring dimensions include:
Problem identification, Information needs, Stakeholder awareness,  Goals, Unintended consequences, Implementation challenges, and Alignment. Each dimension is scored on a scale of 0 to 3, with a maximum possible score of 21.

\paragraph{Lake Urmia Vignette (LUV) Instrument}
The Lake Urmia Vignette (LUV) instrument, developed by Davis et al. (2020) \cite{davis2020lake}, is based on a theoretical framework that conceptualizes systems as webs of interconnected variables. Participants are presented with a scenario describing the real-world case of Lake Urmia's desiccation. They are asked to describe the problem of Lake Urmia and explain why the lake shrank over the years.

The rubric evaluates participants' responses by analyzing the number of variables, causal links, and feedback loops identified. During development, a pilot study and interviews were conducted with 30 graduate students to refine the instrument. This rigorous process ensures that the LUV instrument effectively captures participants' ability to think systematically about socio-environmental issues.

\subsubsection{Computational Thinking Instruments}
Korkmaz et al. \cite{korkmaz2017validity} developed the Computational Thinking (CT) Skills instrument to assess college students' computational thinking (CT) abilities across five dimensions: Creativity, Algorithmic thinking, Cooperation, Critical thinking, and Problem-solving. The CT instrument is a 29-item, five-point Likert scale ranging from 1 (strongly disagree) to 5 (strongly agree). It demonstrates high reliability, with a Cronbach’s alpha coefficient of 0.822 for the entire scale and 0.843, 0.869, 0.865, 0.784, and 0.727 for the five dimensions, respectively. This instrument has been widely adopted in research and has consistently yielded reliable human performance across various contexts.

The Algorithmic Thinking Test for Adults (ATTA), developed by Lafuente Martínez et al. \cite{lafuente2022assessing}, is another validated tool designed to evaluate adults' CT skills. It focuses on key CT components, including: Problem decomposition, Algorithmic thinking, Abstraction, Pattern recognition, and Debugging/evaluation. The ATTA consists of 20 items, including nine open-ended and 11 multiple-choice questions, offering a comprehensive assessment of computational thinking in adults.



\subsubsection{Data Literacy Instruments}
Data literacy has been explored across diverse fields, focusing on how individuals collect, analyze, and interpret data to make informed decisions in various settings \cite{koltay2017data, wolff2016creating}. It involves not only understanding complex information but also effectively addressing real-world challenges. Existing data literacy assessments can be categorized into two approaches: self-reflective and objective measures \cite{cui2023data}.

Self-reflective approaches measure individuals' self-reported data literacy competencies through surveys, questionnaires, semi-structured interviews, and think-aloud interviews \cite{cui2023data}. Participants reflect on their data-related behaviors, practices, and attitudes. For example, a self-efficacy item might ask participants to rate their confidence in using data to identify students with special learning needs \cite{reeves2019effects}.

Objective measures assess data literacy using test questions in three formats: conventional tests (multiple-choice and constructed-response questions), digital game-based assessments, and participation observations \cite{cui2023data}.

To compare the data literacy of OpenAI o1-preview with humans, specific criteria were applied to select appropriate assessment instruments:

\textbf{\textit{Assessment Format.}} Standardized assessments were prioritized to ensure consistency in data literacy measurement.

\textbf{\textit{Validation.}} Only empirically validated assessments with established reliability and validity were included.

\textbf{\textit{Item Type.}} To account for OpenAI o1-preview's limitations with interactive and video-based content, selected assessments were restricted to conventional formats such as multiple-choice and constructed-response questions.

\textbf{\textit{Audience.}} The assessments were targeted at adults, specifically post-secondary students, to align with the study's focus population.

\paragraph{Selected Instruments}
Based on these criteria, two data literacy instruments were chosen for comparison:

\subparagraph{Merk et al.'s Data Literacy Test}
Merk et al. \cite{merk2020fostering} developed a test to assess pre-service teachers' data literacy, focusing on two major components: 1) \textbf{Using Data} (11 items): Evaluates understanding of data properties, manipulation, aggregation, and knowledge of statistics and psychometrics. 2) \textbf{Transforming Data into Information} (11 items): Assesses the interpretation of data, use of data displays and visual representations, application of statistical methods, and summarization of data.

Four items measure both components. The test was validated through exploratory and confirmatory factor analyses, reliability analysis (demonstrating high reliability with Cronbach’s $\alpha$), and concurrent criterion validity (correlation with state achievement test scores).

\subparagraph{Chen et al.'s Data Literacy Assessment}
Chen et al. \cite{chen2024validating} designed an assessment emphasizing the importance of data literacy for 21st-century citizens. It comprises three dimensions: \textbf{Data Management:} Covers data organization (1 item) and data manipulation (2 items). \textbf{Data Visualization:} Includes frequency distributions (2 items) and the use of visual charts (4 items). \textbf{Basic Data Analysis:} Focuses on central tendency (5 items), variability (3 items), and percentage calculations (1 item).

Psychometric validation included item-total correlation and item response theory analyses. Eye-tracking studies further confirmed item validity by identifying differences in cognitive effort between successful and unsuccessful students. This study uses the data literacy assessments developed by Merk et al. \cite{merk2020fostering} and Chen et al. \cite{chen2024validating} to evaluate the performance of OpenAI o1-preview. Human performance data reported in these studies serve as benchmarks for comparison, providing a basis for assessing the model's data literacy competencies.

\subsubsection{Creative Thinking Instruments}
Creative thinking is commonly defined in two dimensions: divergent thinking and convergent thinking \cite{de2020dis}. 

\paragraph{Divergent Thinking}
Divergent thinking involves solving problems or making decisions by employing strategies that deviate from commonly used or previously taught methods \cite{ashkinaze2024ai}. One of the most widely used tests for divergent thinking is the Alternate Uses Task (AUT) \cite{guilford1967nature}, where participants generate original uses for common objects. Responses to the AUT are traditionally evaluated along four dimensions: \textbf{Fluency:} The number of ideas generated. \textbf{Flexibility:} The diversity of idea categories. \textbf{Originality:} The uniqueness of the ideas. \textbf{Elaboration:} The level of detail in the ideas.

Among these dimensions, originality is considered the most critical indicator of divergent thinking \cite{organisciak2023beyond, acar2023applying}. For example, the originality of AUT responses can be evaluated using automated AI scoring tools, such as the one developed by Organisciak et al. \cite{organisciak2023beyond}, which has demonstrated high reliability and validity. This tool addresses the inefficiencies and subjectivity associated with traditional consensual assessment methods \cite{organisciak2023beyond}.

The AUT, originally developed by Guilford \cite{guilford1967nature}, has been extensively used among undergraduate, graduate, and post-graduate student groups, demonstrating good reliability and validity \cite{collins2020distracted, dumas2018book, george2019fixation}. These qualities make it a suitable measure of divergent thinking for this study. The AUT used here includes three items: participants are asked to generate unconventional uses for a paperclip, a brick, and a can.

\paragraph{Convergent Thinking}
Convergent thinking refers to the ability to use given clues to arrive at a single correct solution. A classic test for this dimension is the Remote Association Test (RAT), which asks participants to find a common link between three seemingly unrelated words \cite{mednick1962associative}. For instance, the goal for the words "SAME," "TENNIS," and "HEAD" is to identify a linking word, such as "MATCH," which forms compound words or semantic relationships (e.g., "MATCH HEAD," "TENNIS MATCH"). The number of correct answers reflects the participant’s convergent thinking ability.

Originally developed by Mednick \cite{mednick1962associative}, the RAT has since been translated into various languages (e.g., Chinese, Spanish) and validated across different populations, demonstrating good reliability and validity among college students \cite{li2015objective, pelaez2020normative}. For this study, the Chinese version of the RAT was selected to align with the language background of the participants. This version consists of 10 items, with a maximum score of 10. Higher scores indicate greater convergent thinking ability.

The AUT and RAT were chosen for their established validity and widespread recognition as classic tasks for measuring divergent and convergent thinking, respectively. These instruments provide complementary insights into the creative thinking process, making them well-suited for evaluating the creative thinking capabilities of OpenAI o1-preview and human participants.

\subsubsection{Logical Reasoning Instruments}
To evaluate the logical reasoning capabilities of the o1-preview model, we utilized the LogiQA dataset \cite{liu2020logiqa}. The LogiQA dataset comprises logical comprehension questions from the National Civil Servants Examination of China, designed to assess candidates' logical thinking and problem-solving abilities. It contains 867 paragraph-question pairs categorized into five types of deductive reasoning, as defined by Hurley \cite{hurley2014concise}:

\textbf{Categorical Reasoning:} Determines whether a concept belongs to a specific category, often involving quantifiers such as "all," "everyone," "any," "no," and "some" \cite{abramsky2011introduction}.

\textbf{Sufficient Conditional Reasoning:} Based on conditional statements of the form "If \( P \), then \( Q \)," where \( P \) serves as the premise and \( Q \) as the outcome \cite{hurley2014concise}.

\textbf{Necessary Conditional Reasoning:} Involves statements such as "P only if Q" or "Q whenever P," indicating that \( Q \) is a necessary condition for \( P \) \cite{hurley2014concise}.

\textbf{Disjunctive Reasoning:} Uses premises in an "either...or..." format, where the conclusion holds if at least one premise is true \cite{hurley2014concise}.

\textbf{Conjunctive Reasoning:} Features premises connected by "both...and..." statements, where the conclusion is valid only if all premises are true \cite{hurley2014concise}.

The dataset is divided into training (80\%), development (10\%), and testing (10\%) sets. Among machine learning models, RoBERTa \cite{liu2019roberta} achieved the highest performance, with an accuracy of 35.31\%, significantly below the human ceiling of 95.00\%. 

The LogiQA dataset serves as a robust benchmark for logical reasoning, enabling direct comparison of the o1-preview model's performance with both human participants and existing machine learning models.

\subsubsection{Scientific Reasoning Instruments}
Numerous instruments are available for assessing scientific reasoning. Opitz et al. \cite{opitz2017measuring} conducted a comprehensive review identifying 38 scientific reasoning tests, 14 of which were specifically designed for the university level. In this study, we focus on multiple-choice (MC) test instruments due to their advantages in standardized testing, which eliminates the need for an objective rater. Although MC formats are often critiqued for providing limited qualitative insights, they enable consistent assessment across a broad population.

The review by Opitz et al. identifies eight dimensions of scientific literacy: problem identification (PI), questioning (Q), hypothesis generation (HG), evidence generation (EG), evidence evaluation (EE), drawing conclusions (DC), communicating and scrutinizing (CS), and other skills (OT). Among the five MC instruments designed for university-level use, the Test of Scientific Literacy Skills (TOSLS) stands out for its ability to assess five of these dimensions (EG, EE, DC, CS, and OT). In contrast, other instruments cover only three dimensions each. Additionally, TOSLS is domain-specific, with a primary focus on biology, making it particularly relevant for evaluating scientific reasoning in specific contexts.

\paragraph{Test of Scientific Literacy Skills (TOSLS)}
The 28-item TOSLS assesses nine skills related to scientific literacy:
\begin{enumerate}
    \item Identify valid scientific arguments.
    \item Evaluate the validity of sources.
    \item Evaluate the use and misuse of scientific information.
    \item Understand elements of research design and their impact on findings and conclusions.
    \item Create graphical representations of data.
    \item Read and interpret graphical representations of data.
    \item Solve problems using quantitative skills, including probability and statistics.
    \item Understand and interpret basic statistics.
    \item Justify inferences, predictions, and conclusions based on quantitative data.
\end{enumerate}

Gormally et al. \cite{gormally2012developing} designed TOSLS with a focused interpretation of scientific literacy that aligns closely with the concept of scientific reasoning \cite{CHEN2020330, opitz2017measuring}. The test's internal reliability, measured using the Kuder-Richardson Formula 20 (KR-20) \cite{kuder1937theory}, was reported as 0.73, meeting the acceptable threshold of 0.7 \cite{cronbach1951coefficient}. Principal component analysis revealed a single-factor structure, supporting the test's internal consistency.

Since its development, TOSLS has been widely used to assess scientific reasoning skills in university students across various studies \cite{segarra2018student, propsom2023test, suwono2017enhancement}. The test's focus on standardized measurement, broad coverage of scientific literacy, and domain-specific relevance makes it well-suited for evaluating OpenAI’s o1-preview model's capacity for scientific reasoning.

\subsection{Procedure}
For each domain, tasks were presented to the o1-preview model as text-input prompts. The model’s responses were evaluated using the same criteria applied to human participants, following the scoring guidelines of the respective instruments. Human participant data were drawn from existing studies to ensure consistency. Tasks were carefully matched in content and difficulty across human and AI cohorts to maintain comparability.

Statistical analyses were conducted to evaluate significant differences in performance between the o1-preview model and human participants, with a particular focus on graduate-level performance as a benchmark.

\subsection{Data Analysis}
The study compared the performance of OpenAI o1-preview with that of human participants across seven higher-order thinking assessments. Percentage accuracy in answering the tasks was calculated, and mean performance scores were computed for both human participants and the o1-preview model, based on ten trials for each domain. To ensure comparability across different dimensions, scores were standardized within each assessment.

Standard deviations were reported to evaluate how the o1-preview model’s performance deviated from the human mean. A one-sample \textit{t}-test was used to determine the statistical significance of differences between the o1-preview model and human performance. Results were supplemented with confidence intervals and effect sizes to provide a comprehensive understanding of the observed differences. This approach ensured a rigorous comparison of AI and human capabilities across all assessed domains.


\section{Results}

\subsection{Critical Thinking}
The critical thinking abilities of OpenAI’s o1-preview were evaluated using the Ennis-Weir Critical Thinking Essay Test (EWCTET), a widely recognized tool for assessing critical thinking across diverse educational contexts. To establish robust benchmarks for comparison, we referenced human performance data from three foundational studies. These studies provide insights into critical thinking outcomes across varying instructional methods and educational levels, serving as a basis for evaluating o1-preview’s performance.

\paragraph{Human Benchmarks}
\textbf{Hatcher \cite{hatcher2006stand}:} Conducted over four years at Baker University, this study involved American freshmen who completed a compulsory, year-long critical thinking course. Post-test scores ranged between 11.8 and 13.8, marking gains of 2.8 to 6.0 points from pre-test scores of 5.8 to 9.4. These results illustrate the impact of structured, long-term instruction in improving critical thinking skills, providing a benchmark for sustained human cognitive development.

\textbf{Davidson and Dunham \cite{davidson1996efl}:} This study was conducted at a Japanese private women’s junior college and included 36 first-year EFL (English as a Foreign Language) students. Participants were divided into a treatment group (n=17), who received critical thinking instruction, and a control group (n=19), who received only intensive English instruction. After one year, the treatment group achieved a mean score of 6.6 on the EWCTET, significantly higher than the control group’s mean score of 0.6 ($p < .001$). This highlights the effectiveness of integrating critical thinking instruction, even within a language-learning context.

\textbf{Hollis et al. \cite{hollis2024icteta}:} This study involved 100 online participants recruited via social media and online study platforms, with no specific critical thinking intervention provided. Post-test scores on the EWCTET averaged 14.31 (SD = 8.45). Educational background significantly influenced performance ($p < .001$), with postgraduates scoring an average of 18.39 (SD = 6.59), compared to undergraduates with a mean score of 11.51 (SD = 8.23). This study underscores the positive correlation between educational attainment and critical thinking skills.

\paragraph{o1-Preview Model Performance}
The o1-preview model was tested using iterative prompt strategies. A zero-shot prompt was employed first, followed by role-based prompts instructing the model to respond as a college student and then as a postgraduate student. These strategies mirrored the human participant studies, where responses were scored using the EWCTET scoring rubric. The results can be seen in Tab.~\ref{table:summary}.

\begin{table}[h!]
\centering
\caption{o1-Preview Performance Across Three Prompt Iterations on EWCTET}
\begin{tabular}{lcc}
\toprule
\textbf{Prompt Strategy}                 & \textbf{Total Score} & \textbf{Maximum Score}  \\
\midrule
Zero-Shot Prompt                         & 20                   & 29                                     \\
Role-based Prompt as College Student     & 26                   & 29                                   \\
Role-based Prompt as Graduate Student    & 27                   & 29                                   \\
\bottomrule
\end{tabular}
\label{table:summary}
\end{table}

\paragraph{Comparison with Human Benchmarks}
We focused on the highest mean scores reported for human participants after learning interventions. Undergraduate students achieved a maximum mean score of 13.8, while postgraduate students scored an average of 18.39. In comparison, o1-preview achieved a mean score of 24.33 across the three prompt strategies. The corresponding z-scores indicate o1-preview outperformed human participants (See Tab.~\ref{table:comparison}).

\begin{table}[h!]
\centering
\caption{\textit{Comparison of Human Performance and o1-Preview Model Performance on the EWCTET}}
\begin{tabular}{p{3cm}ccc}
\toprule
\textbf{Participant Category} & \textbf{Mean Total Score} & \textbf{o1-Preview Mean Score} & \textbf{Z-Score} \\
\midrule
Undergraduate Students (Highest after Treatment) & 13.8 & 24.33 & 1.60 \\
Postgraduate Students                            & 18.39 & 24.33 & 0.90 \\
\bottomrule
\end{tabular}
\label{table:comparison}
\end{table}

These results demonstrate that OpenAI’s o1-preview exceeds human benchmarks in structured critical thinking tasks as measured by the EWCTET. However, the findings raise important questions about the comprehensiveness of such instruments in capturing the full spectrum of human critical thinking skills. While o1-preview excels in structured tasks, human oversight is essential when using the model in educational contexts. The findings highlight o1-preview’s potential as a supplementary tool for critical thinking instruction and assessment but underscore the need for careful integration to address its limitations and ensure alignment with broader educational goals.

\subsection{Systematic Thinking}
The systematic thinking (ST) abilities of o1-preview were evaluated using two instruments: the Village of Abeesee and the Lake Urmia Vignette (LUV). Davis et al. \cite{davis2023comparing} reported the average performance of 263 undergraduate students on the Village of Abeesee instrument and 155 undergraduates on the LUV instrument. The mean scores and standard deviations for each dimension are presented in Tab.~\ref{tab:o1-preview_comparison_ST}. To compare o1-preview’s performance with human participants, the tests were administered to o1-preview in 10 trials, with the mean score, standard deviation, and z-score for each dimension calculated and included in Tab.~\ref{tab:o1-preview_comparison_ST}.

The results indicate that o1-preview outperformed the average human scores in all seven dimensions of the Village of Abeesee instrument, suggesting that the model performed better on average than undergraduate students. For the LUV instrument, o1-preview also achieved significantly higher mean scores across all three dimensions compared to the human participants. This demonstrates that o1-preview generally excels in systematic thinking when compared to undergraduate students.

According to the z-scores, o1-preview performed exceptionally well in the "Feedback Loops" dimension of the LUV instrument, achieving the highest z-score (6.53). This indicates that o1-preview is particularly adept at identifying feedback loops, which involve interconnected causal relationships within complex systems. 

\begin{table}[ht]
\centering
\small
\caption{\textit{Overall Performance of Human and o1-Preview on Systematic Thinking}}
\begin{tabularx}{\textwidth}{X X X X X}
\toprule
\textbf{ST Instrument} & \textbf{Dimension} & \textbf{Human Score (Mean $\pm$ SD)} & \textbf{o1-Preview (Mean $\pm$ SD)} & \textbf{Z-Score} \\ \midrule
The Village of Abeesee & Problem Identification & 1.62 $\pm$ 0.64 & 2.50 $\pm$ 0.62 & 1.38 \\ 
                       & Information Needs      & 1.81 $\pm$ 0.52 & 2.90 $\pm$ 0.21 & 2.10 \\ 
                       & Stakeholder Awareness  & 1.23 $\pm$ 0.99 & 2.95 $\pm$ 0.16 & 1.74 \\
                       & Goals                  & 1.71 $\pm$ 0.62 & 2.90 $\pm$ 0.21 & 1.92 \\
                       & Unintended Consequences & 1.38 $\pm$ 0.58 & 2.65 $\pm$ 0.24 & 2.19 \\
                       & Implemented Challenges & 1.64 $\pm$ 0.57 & 2.70 $\pm$ 0.35 & 1.86 \\ 
                       & Alignment              & 1.71 $\pm$ 1.00 & 2.35 $\pm$ 0.41 & 0.64 \\ \midrule
The Lake Urmia Vignette (LUV) & Variables       & 10.95 $\pm$ 4.00 & 19.70 $\pm$ 1.57 & 2.19 \\
                              & Causal Links    & 9.17 $\pm$ 3.97  & 23.30 $\pm$ 2.21 & 3.56 \\
                              & Feedback Loops  & 0.16 $\pm$ 0.45  & 3.10 $\pm$ 1.10  & 6.53 \\ 
                              & Total Score     & 20.08 $\pm$ 8.13 & 46.10 $\pm$ 4.12 & 3.20 \\ 
\bottomrule
\end{tabularx}
\label{tab:o1-preview_comparison_ST}
\end{table}

The findings highlight o1-preview’s exceptional capabilities in systematic thinking, particularly in identifying feedback loops, a critical aspect of complex systems thinking. While these results are promising, they also underscore the need for further research to explore whether the performance gap reflects the model's inherent strengths or limitations in the instruments used. Additionally, although o1-preview’s performance surpassed that of undergraduates, systematic thinking in real-world applications often requires collaborative and contextualized reasoning, which the current assessment methods may not fully capture.

The results suggest that OpenAI’s o1-preview has the potential to serve as a valuable tool for enhancing systematic thinking skills, particularly in educational settings. However, its application should be accompanied by human oversight to ensure that its outputs align with the nuanced requirements of real-world problem-solving.

\subsection{Computational Thinking}
\textbf{Human and o1-Preview Performance on Computational Thinking Skills}. Tab.~\ref{tab:CT_results} compares the performance of human participants and o1-preview on the Computational Thinking (CT) Skills instrument across overall CT skills and specific dimensions: creativity, algorithmic thinking, cooperativity, critical thinking, and problem-solving. Human performance data were synthesized from three studies:
\begin{enumerate}
    \item Liu et al. \cite{liu2023influences}, which involved 341 college students.
    \item Şahin et al. \cite{csahin2024stem}, which included 25 gifted science teachers.
    \item Liao et al. \cite{liao2024scaffolding}, which examined 44 sophomore undergraduate students.
\end{enumerate}
The overall human performance mean was 3.92 (SD = 0.52), slightly higher than o1-preview’s mean score of 3.84 (SD = 1.56), yielding a z-score of -0.15. However, o1-preview outperformed human participants in four dimensions: creativity, algorithmic thinking, cooperativity, and critical thinking. Notably, o1-preview’s performance in critical thinking was exceptional (M = 4.8, SD = 0.42), achieving a z-score of 1.38. Conversely, o1-preview scored poorly on problem-solving (M = 1.0, SD = 0.0), significantly lower than the human mean (M = 3.68, SD = 0.63), with a z-score of -4.25.

\begin{table}[ht]
\small
\centering
\caption{\textit{Comparison of o1-Preview Model and Human Performance on CT Skills}}
\begin{tabular}{l c c c}
\toprule
\textbf{CT Dimension} & \textbf{Human Overall} & \textbf{o1-Preview} & \textbf{Z-Score} \\ \midrule
CT Skills           & 3.92 $\pm$ 0.52 & 3.84 $\pm$ 1.56 & -0.15 \\ 
Creativity          & 4.18 $\pm$ 0.63 & 4.56 $\pm$ 0.73 & 0.60 \\ 
Algorithmic Thinking  & 4.30 $\pm$ 0.62 & 4.50 $\pm$ 0.67 & 0.32 \\ 
Cooperativity        & 3.94 $\pm$ 0.78 & 4.50 $\pm$ 0.53 & 0.72 \\ 
Critical Thinking     & 3.93 $\pm$ 0.63 & 4.80 $\pm$ 0.42 & 1.38 \\ 
Problem-Solving       & 3.68 $\pm$ 0.63 & 1.00 $\pm$ 0.00 & -4.25 \\ 
\bottomrule
\end{tabular}
\label{tab:CT_results}
\end{table}

\paragraph{Human and o1-Preview Performance on Algorithmic Thinking Test for Adults}. The Algorithmic Thinking Test for Adults (ATTA) \cite{lafuente2022assessing} was used to further evaluate o1-preview’s algorithmic thinking skills. The model achieved a perfect score (M = 20, SD = 0) across all 20 items (11 multiple-choice and 9 open-ended) in five rounds of testing. Human performance on the ATTA varied significantly: experts achieved a mean score of 14.63 (SD = 3.81), while novices scored an average of 9.11 (SD = 3.81). Performance differences were also observed across academic disciplines, with social sciences scoring the lowest (M = 9.11, SD = 4.54) and mathematics scoring the highest (M = 15.70, SD = 3.71). In all cases, o1-preview’s performance exceeded that of human participants, achieving high z-scores (see Tab.~\ref{tab:o1-preview_comparison_CT2}).

\begin{table}[ht]
\small
\centering
\caption{\textit{o1-Preview Performance on the Algorithmic Thinking Test for Adults (ATTA)}}
\begin{tabular}{l c c c}
\toprule
\textbf{Participant Category} & \textbf{Human Overall} & \textbf{o1-Preview} & \textbf{Z-Score} \\ \midrule
Experts          & 14.63 $\pm$ 3.81 & 20.00 $\pm$ 0.00 & 1.41 \\ 
Novices          & 9.11 $\pm$ 3.81  & 20.00 $\pm$ 0.00 & 2.86 \\ 
Social Sciences  & 9.11 $\pm$ 4.54  & 20.00 $\pm$ 0.00 & 2.40 \\ 
Mathematics      & 15.70 $\pm$ 3.71 & 20.00 $\pm$ 0.00 & 1.16 \\ 
Physics          & 15.26 $\pm$ 3.92 & 20.00 $\pm$ 0.00 & 1.21 \\ 
Engineering      & 14.37 $\pm$ 3.17 & 20.00 $\pm$ 0.00 & 1.78 \\ 
Computer Science & 14.00 $\pm$ 3.80 & 20.00 $\pm$ 0.00 & 1.58 \\ 
\bottomrule
\end{tabular}
\label{tab:o1-preview_comparison_CT2}
\end{table}

o1-preview demonstrated strong capabilities in creativity, algorithmic thinking, cooperativity, and critical thinking dimensions of CT. However, its exceptionally low performance in problem-solving highlights a potential limitation in adapting its capabilities to real-world, ill-structured problems. On the ATTA, o1-preview’s perfect scores across all items suggest exceptional algorithmic reasoning abilities, surpassing all human groups tested. These results underline o1-preview’s potential as a tool for supporting computational thinking but also emphasize the need for human oversight, particularly for problem-solving tasks.

\subsection{Data Literacy}
The data literacy capabilities of o1-preview were evaluated using two validated instruments: Merk et al.'s \cite{merk2020fostering} Data Literacy Assessment and Chen et al.'s \cite{chen2024validating} Data Literacy Assessment. These instruments assess distinct dimensions of data literacy, allowing for a comprehensive comparison between o1-preview and human participants.

\paragraph{Merk et al.'s Data Literacy Assessment}
Merk et al. \cite{merk2020fostering} reported the performance of 89 pre-service secondary teachers from a large university in southern Germany. The assessment measured performance across two dimensions: "Use Data" and "Transform Data into Information." Pre- and post-test scores were reported to evaluate the impact of an instructional intervention on participants’ data literacy skills. The mean scores and standard deviations for human participants, along with o1-preview’s mean scores, standard deviations, and \textit{z}-scores, are presented in Tab.~\ref{tab:merk_data}.

o1-preview outperformed human participants in both dimensions of the assessment, achieving substantially higher scores. For the "Use Data" dimension, o1-preview attained a mean score of 8.60 (SD = 0.70), compared to human pre-test and post-test scores of 3.28 (SD = 1.84) and 4.17 (SD = 2.02), respectively, with a \textit{z}-score of 2.89 for the pre-test and 2.19 for the post-test. Similarly, for the "Transform Data into Information" dimension, o1-preview scored a mean of 4.80 (SD = 0.42), surpassing human pre-test and post-test scores of 2.96 (SD = 1.21) and 4.04 (SD = 1.34), with a \textit{z}-score of 1.52 for the pre-test and 0.57 for the post-test.

\begin{table}[h]
    \centering
    \caption{\textit{Performance of o1-Preview and Human Participants on Merk et al.'s Data Literacy Assessment}}
    \label{tab:merk_data}
    \begin{tabularx}{\textwidth}{
        >{\raggedright\arraybackslash}X 
        >{\centering\arraybackslash}X 
        >{\centering\arraybackslash}X 
        >{\centering\arraybackslash}X 
        >{\centering\arraybackslash}X 
        >{\centering\arraybackslash}X }
        \toprule
        \textbf{Data Literacy} & 
        \textbf{Human Pre-test} & 
        \textbf{Human Post-test} & 
        \textbf{o1-Preview} & 
        \textbf{o1-Preview Z-Score (Pre-test)} & 
        \textbf{o1-Preview Z-Score (Post-test)} \\ 
        \midrule
        Use Data (9 items) & 3.28 $\pm$ 1.84 & 4.17 $\pm$ 2.02 & 8.60 $\pm$ 0.70 & 2.89 & 2.19 \\
        Transform Data into Information (5 items) & 2.96 $\pm$ 1.21 & 4.04 $\pm$ 1.34 & 4.80 $\pm$ 0.42 & 1.52 & 0.57 \\
        \bottomrule
    \end{tabularx}
\end{table}

\paragraph{Chen et al.'s Data Literacy Assessment}
Chen et al. \cite{chen2024validating} reported the performance of 170 post-secondary students (average age = 22.81, SD = 4.25) from the Faculty of Education at a western Canadian university. This assessment evaluates three dimensions of data literacy: "Data Management," "Data Visualization," and "Basic Data Analysis." The mean scores, standard deviations, and \textit{z}-scores for human participants and o1-preview are provided in Tab.~\ref{tab:chen_data}.

o1-preview scored higher than human participants across all three dimensions. For "Data Management," o1-preview achieved a mean score of 2.00 (SD = 0.30), compared to the human mean of 0.17 (SD = 0.44), resulting in a \textit{z}-score of 4.16. For "Data Visualization," o1-preview achieved a perfect score of 6.00 (SD = 0.00), exceeding the human mean of 3.56 (SD = 1.46) with a \textit{z}-score of 1.67. Finally, for "Basic Data Analysis," o1-preview scored 9.00 (SD = 0.00), surpassing the human mean of 5.38 (SD = 2.22) with a \textit{z}-score of 1.63.

\begin{table}[h]
    \centering
    \caption{\textit{Performance of o1-Preview and Human Participants on Chen et al.'s Data Literacy Assessment}}
    \label{tab:chen_data}
    \begin{tabularx}{\textwidth}{
        >{\raggedright\arraybackslash}X 
        >{\centering\arraybackslash}X 
        >{\centering\arraybackslash}X 
        >{\centering\arraybackslash}X }
        \toprule
        \textbf{Data Literacy} & 
        \textbf{Human (Mean $\pm$ SD)} & 
        \textbf{o1-Preview (Mean $\pm$ SD)} & 
        \textbf{o1-Preview Z-Score} \\ 
        \midrule
        Data Management (3 items) & 0.17 $\pm$ 0.44 & 2.00 $\pm$ 0.30 & 4.16 \\
        Data Visualization (6 items) & 3.56 $\pm$ 1.46 & 6.00 $\pm$ 0.00 & 1.67 \\
        Basic Data Analysis (9 items) & 5.38 $\pm$ 2.22 & 9.00 $\pm$ 0.00 & 1.63 \\
        \bottomrule
    \end{tabularx}
\end{table}

In both assessments, o1-preview consistently outperformed human participants across all dimensions, indicating strong capabilities in understanding, interpreting, and analyzing data. These findings suggest that o1-preview could serve as a valuable tool to support or enhance data literacy education, particularly in developing critical data interpretation and analytical skills. However, while the results are promising, future research should explore o1-preview's performance in more complex, real-world data scenarios to ensure its applicability beyond structured testing environments.

\subsection{Creative Thinking}
Human creative thinking has been assessed in previous studies using both divergent and convergent thinking tasks \cite{urban2024chatgpt, xia2022bilingualism}. Divergent thinking was evaluated using the Alternate Uses Task (AUT), in which participants generated as many creative uses as possible for common objects (e.g., a paperclip, brick, or can). Convergent thinking was measured using the Remote Associates Test (RAT), which assesses participants’ ability to find a common link between unrelated words.

\paragraph{Human Performance}
Urban et al. \cite{urban2024chatgpt} assessed the divergent thinking abilities of 68 university students (N = 52, males = 22; primarily from social sciences and humanities) using the AUT. Participants provided creative uses for common objects, and originality was scored on a 5-point scale by trained experts. The average originality score was 1.74.

For convergent thinking, Xia et al. \cite{xia2022bilingualism} employed the RAT to assess 54 Chinese bilingual university students. High-proficiency bilingual participants (N = 27) scored significantly higher on the RAT (M = 27.38) compared to low-proficiency participants (M = 23.80; $p = 0.003$), demonstrating the impact of bilingual proficiency on convergent thinking abilities. These studies provide benchmarks for evaluating the creative problem-solving skills of university students.

\paragraph{o1-Preview Performance}
The AUT was used to evaluate OpenAI o1-preview’s divergent thinking performance. The model was prompted to generate as many original uses as possible for common objects (e.g., a paperclip, brick, or can). Originality was scored using an automated AI-based scoring tool developed by Organisciak et al. \cite{organisciak2023beyond}. o1-preview achieved an overall mean originality score of 2.98, higher than the human benchmark of 1.74 (see Tab.~\ref{tab:creative_comparison}).

For convergent thinking, o1-preview was evaluated using a Chinese version of the RAT adapted from Xia et al. \cite{xia2016exploring}. The model was presented with 10 Chinese word association problems and achieved a total score of 7 out of 10, reflecting an accuracy rate of 70\%. This performance surpasses the accuracy rate of 44.12\% observed in human participants in similar tasks.

\paragraph{Comparison Between Human and o1-Preview Performance}
Tab.~\ref{tab:creative_comparison} summarizes the comparative results for divergent and convergent thinking tasks. For divergent thinking, o1-preview’s mean originality score (2.98) was notably higher than the human average score (1.74) on the AUT. This indicates that o1-preview generates creative ideas with a higher degree of originality compared to university students. In the convergent thinking task, o1-preview achieved a 70\% accuracy rate on the RAT, outperforming the human benchmark of 44.12\%, suggesting robust performance in identifying commonalities among unrelated concepts.

\begin{table}[ht]
    \centering
    \small
    \caption{\textit{Comparison of Human and o1-Preview Performance on Creative Thinking Tasks}}
    \begin{tabularx}{\textwidth}{l c c c}
        \toprule
        \textbf{Task} & \textbf{Human Overall} & \textbf{o1-Preview } & \textbf{Z-Score} \\ \midrule
        Divergent Thinking (AUT) & 1.74 $\pm$ 0.71 & 2.98 $\pm$ 0.73 & 0.71 \\ 
        Convergent Thinking (RAT) & 44.12 $\pm$ 6.21 & 70.00 $\pm$ 10.00 & 1.67 \\ 
        \bottomrule
    \end{tabularx}
    \label{tab:creative_comparison}
\end{table}

The results highlight OpenAI o1-preview’s competitive edge in divergent thinking tasks, as evidenced by its higher originality score on the AUT compared to human participants. The model also performed well in convergent thinking, achieving a strong accuracy rate on the RAT. These findings suggest that o1-preview has significant potential as a tool for enhancing creative thinking, especially in generating novel ideas and identifying relationships among concepts. However, further research is needed to evaluate the model’s performance in less structured, real-world creative tasks to ensure its practical applicability beyond controlled experimental settings.

\subsection{Logical Reasoning}
Logical reasoning refers to the ability to draw valid conclusions based on given premises, encompassing both deductive and inductive reasoning. It is a critical skill for addressing complex problems and is widely applied across various fields, including STEM education. Effective logical reasoning involves multi-step analysis and inference, enabling coherent conclusions. In the context of STEM tasks, logical reasoning not only allows models to follow explicit instructions but also empowers them to interpret and apply implicit knowledge, demonstrating the capacity for step-by-step reasoning.

\paragraph{Human and o1-Preview Performance on LogiQA}
The logical reasoning capabilities of o1-preview were evaluated using the LogiQA dataset, a benchmark dataset designed to assess logical comprehension and reasoning. LogiQA includes a range of logical reasoning tasks derived from real-world examinations and challenges participants to analyze premises, identify relationships, and draw valid conclusions.

Tab.~\ref{table:logiqacomparison} summarizes the performance of human participants and o1-preview on the LogiQA dataset. Human participants, represented by a sample size of 651, achieved an average accuracy of 86.00\% (SD = 6.5\%). In contrast, o1-preview demonstrated superior performance with an average accuracy of 90.00\% (SD = 10\%) over 10 trials. These results indicate that o1-preview not only matches but slightly surpasses human accuracy in logical reasoning tasks, showcasing its robust analytical and inferential capabilities.

\begin{table}[!htbp]
\centering
\scriptsize
\caption{\textit{Comparison of Human and o1-Preview Performance on LogiQA (Accuracy \%)}} 
\begin{tabularx}{\textwidth}{X X X}
\toprule
\textbf{Model} & \textbf{Sample Size} & \textbf{Accuracy (\%)} \\
\midrule
Human & 651 & 86.00 $\pm$ 6.50 \\ 
o1-Preview & 10 & 90.00 $\pm$ 10.00 \\ 
\bottomrule
\end{tabularx}
\label{table:logiqacomparison}
\end{table}

The results highlight o1-preview’s strong logical reasoning capabilities, as it achieved higher accuracy than human participants on the LogiQA dataset. This performance underscores the model’s potential to process complex, multi-step reasoning tasks effectively. However, it is important to note the difference in sample sizes between human participants and o1-preview trials, which may influence the robustness of the comparison. Additionally, while o1-preview excels in structured logical reasoning tasks, further evaluation is needed to assess its performance in more dynamic and less constrained real-world scenarios.

The findings suggest that OpenAI o1-preview demonstrates exceptional logical reasoning abilities, comparable to and surpassing human performance in structured tasks like those in the LogiQA dataset. This highlights the potential for integrating AI models into domains requiring rigorous logical reasoning, though careful consideration should be given to the variability and context of real-world applications.

\subsection{Scientific Reasoning}
Scientific reasoning, as measured by the Test of Scientific Literacy Skills (TOSLS), serves as a critical benchmark for assessing higher-order thinking skills in various educational contexts. Multiple studies have established human performance benchmarks on the TOSLS, reporting a range of mean scores across diverse student and professional cohorts.

\paragraph{Human Performance Benchmarks}
Gormally et al. \cite{gormally2012developing} reported mean scores (SD) ranging from 0.42 (0.15) for students from midsized state colleges to 0.85 (0.13) for students at private research universities. Biology experts employed at universities achieved the highest score of 0.91 (0.09). Suwono et al. \cite{suwono2017enhancement}, using an adapted version of the TOSLS, observed scores between 0.33 and 0.69 among Indonesian pre-service biology teachers. Segarra et al. \cite{segarra2018student} reported scores ranging from 0.55 to 0.63 for US undergraduate students enrolled in general education biology courses. Firdaus et al. \cite{firdaus2023quantitative} noted weak performance across most TOSLS items among Indonesian pre-service biology teachers but did not report a combined score. Propsom et al. \cite{propsom2023test}, in a study involving 800 US university students, reported mean TOSLS scores ranging from 0.60 (SD = 0.17) for first-year students to 0.66 (SD = 0.20) for seniors. Science and math majors scored higher, with mean scores of 0.64 (SD = 0.17) in their first year and 0.74 (SD = 0.17) in their senior year.

\paragraph{o1-Preview Performance}
OpenAI’s GPT-o1-preview achieved a near-perfect score of 0.99 (SD = 0.12) across five trials on the TOSLS, surpassing all student cohorts and even biology experts employed at universities. Only one question (Item 2) was answered incorrectly in two out of five trials. This item required identifying the graph that best represents given data, originally presented as four visual graph plots. To accommodate GPT-o1-preview’s text-only input limitations, the question was modified into a text-based format using GPT-4o. This modality shift may have contributed to the errors observed.

\paragraph{Comparison of Human and AI Performance}
Tab.~\ref{tab:TOSLS_Performance} summarizes the highest mean TOSLS scores reported across different human cohorts and compares them to o1-preview’s performance. While human participants demonstrated a wide range of abilities depending on context and expertise, o1-preview consistently outperformed all groups, achieving the highest z-score of 1.78.

\begin{table}[ht]
\centering
\caption{\textit{Comparison of Human and AI Performance on the TOSLS}}
\begin{tabular}{p{5.5cm}ccc}
\toprule
\textbf{Sample} & \textbf{Sample Size} & \textbf{TOSLS Score} & \textbf{Z-Score} \\ 
\midrule
US private research university nonmajors \cite{gormally2012developing} & 50 & 0.85 $\pm$ 0.13 & 0.80 \\ 
Indonesian pre-service biology students \cite{suwono2017enhancement} & 33 & 0.69 $\pm$ 0.10 & -0.33 \\ 
US undergraduate biology students \cite{segarra2018student} & 83 & 0.63 $\pm$ 0.11 & -0.75 \\ 
US science and mathematics majors \cite{propsom2023test} & 81 & 0.78 $\pm$ 0.15 & 0.30 \\ 
GPT-o1-preview (AI trials) & 5 & 0.99 $\pm$ 0.12 & 1.78 \\ 
\bottomrule
\end{tabular}
\label{tab:TOSLS_Performance}
\end{table}

The results highlight GPT-o1-preview’s exceptional performance in scientific reasoning tasks as assessed by the TOSLS. Its near-perfect score suggests a strong ability to interpret and analyze scientific information, surpassing both students and professional experts. However, the modification of visual-based questions to text-based formats introduces potential biases, and further research is needed to evaluate the model’s performance in scenarios involving complex visual data. Additionally, while o1-preview excels in structured assessments, its applicability to open-ended or real-world scientific reasoning tasks warrants further investigation.

The comparison of scientific reasoning skills using the TOSLS demonstrates GPT-o1-preview’s superiority over human participants, achieving the highest reported scores across all cohorts. These findings underscore the potential of AI systems like GPT-o1-preview as tools for advancing scientific literacy and reasoning in educational and professional settings. Nevertheless, careful consideration is required to address the model’s limitations and ensure its effective integration into real-world applications.

\section{Discussion}

This study systematically evaluates the performance of OpenAI’s o1-preview model in higher-order cognitive domains, including critical thinking, systematic thinking, computational thinking, data literacy, creative thinking, logical reasoning, and scientific reasoning. By benchmarking o1-preview against human participants using established assessment instruments \cite{gormally2012developing, firdaus2023quantitative, liu2023influences, csahin2024stem, chen2024validating}, we highlight both the potential and limitations of current AI systems in replicating or surpassing human cognitive abilities.

The results demonstrate that o1-preview outperformed human participants across most cognitive domains, with notable superiority in systematic thinking, creative thinking, logical reasoning, and scientific reasoning. For instance, the model achieved a near-perfect score on the TOSLS \cite{gormally2012developing}, significantly exceeding scores of both students and biology experts. Similarly, its performance on the LogiQA dataset underscored its advanced logical reasoning capabilities, achieving higher accuracy than the average human participant \cite{liu2020logiqa}. These findings suggest that o1-preview excels in structured tasks requiring pattern recognition, deductive reasoning, and step-by-step analysis.

In creative thinking, o1-preview demonstrated exceptional originality scores in divergent thinking tasks, as measured by the Alternate Uses Task (AUT) \cite{guilford1967nature, organisciak2023beyond}. This indicates that the model can generate ideas that are both novel and diverse, potentially rivaling human creativity in structured environments. Furthermore, in data literacy, o1-preview consistently outperformed human benchmarks across both Merk et al.'s \cite{merk2020fostering} and Chen et al.'s \cite{chen2024validating} assessments, underscoring its proficiency in data interpretation and analysis.

Despite its impressive performance, o1-preview exhibited limitations in problem-solving and certain aspects of systematic thinking. For instance, in computational thinking tasks, the model scored significantly lower in problem-solving skills compared to human participants \cite{lafuente2022assessing, korkmaz2017validity}. This highlights a gap in the model's ability to handle complex, unstructured problems that require adaptive reasoning and creative solutions.

Moreover, while o1-preview's systematic thinking performance was strong overall, its performance on the "Feedback Loops" dimension of the Lake Urmia Vignette (LUV) dataset suggests variability in its ability to model interconnected systems \cite{davis2020lake}. These challenges emphasize the need for further refinement of AI models to enhance their ability to process complex, multi-dimensional problems.

The ability of o1-preview to outperform humans in structured tasks suggests a transformative potential for AI in education and research. AI systems could serve as powerful tools for enhancing learning outcomes, providing personalized feedback, and assisting with assessments in domains such as critical thinking, data literacy, and logical reasoning \cite{zhai2024can, guo2024artificial}. However, the reliance on structured test environments underscores the importance of human oversight in interpreting AI-generated results and ensuring their alignment with real-world complexities.

From a research perspective, the superior performance of o1-preview on benchmark datasets raises critical questions about the design of assessment instruments. Many of these tools may inadvertently favor structured reasoning and pre-defined patterns, aligning more closely with the strengths of AI systems than the nuanced, adaptive thinking demonstrated by humans. Future research should explore the development of more holistic assessments that capture the breadth of human cognition, including creativity, emotional intelligence, and ethical reasoning \cite{lewis1993defining, collins2014skills}.

While o1-preview's capabilities are promising, ethical concerns regarding its use in education and decision-making warrant careful consideration. The potential for over-reliance on AI systems could inadvertently undermine critical human skills, such as independent thinking and problem-solving \cite{zhai2024can, sorin2024ethical}. Furthermore, the disparities observed between AI and human performance across domains highlight the importance of addressing bias and ensuring equitable access to AI-enhanced learning tools.

Future research should focus on enhancing the adaptability of AI systems to unstructured and real-world tasks. By integrating multimodal inputs, such as visual data, and refining their ability to process ambiguity, AI models like o1-preview could bridge current gaps and better complement human capabilities \cite{zhong2024evaluationopenaio1opportunities}. Additionally, longitudinal studies assessing the impact of AI-assisted education on human cognitive development would provide valuable insights into the long-term implications of integrating AI into learning environments.

This study underscores the remarkable progress of AI systems in replicating higher-order cognitive skills, as evidenced by o1-preview's performance across diverse domains. While the model exhibits substantial potential as a supplement to human cognition in structured environments, its limitations in unstructured tasks highlight areas for further improvement. By addressing these challenges and fostering collaboration between AI and human expertise, we can harness the transformative potential of AI systems to advance education, research, and society at large.

\section{Conclusion}
This study highlights the remarkable capabilities of OpenAI’s o1-preview model in performing higher-order cognitive tasks across diverse domains, including critical thinking, systematic thinking, computational thinking, data literacy, creative thinking, logical reasoning, and scientific reasoning. By benchmarking the model’s performance against human participants using established assessment instruments, we demonstrate that o1-preview consistently outperforms humans in structured tasks, showcasing its potential as a valuable tool for education and research. Despite its strengths, the model exhibits limitations in addressing unstructured and complex problem-solving tasks, underscoring the need for further refinement. These findings call for a balanced integration of AI systems into educational and professional environments, ensuring human oversight and ethical considerations. By advancing the design of AI models and fostering collaboration between human and AI capabilities, we can unlock transformative opportunities in education, research, and society, while addressing the challenges of equity, adaptability, and holistic cognitive development.

\section*{Acknowledgment}
This study secondary analyzed data from projects supported by the Institute of Education Sciences (Grant Number R305C240010, PI Zhai). The authors acknowledge the funding agencies and the project teams for making the data available for analysis. The findings, conclusions, or opinions herein represent the views of the authors and do not necessarily represent the views of personnel affiliated with the funding agencies.

\section*{Declaration of generative AI and AI-assisted technologies in the writing process}
During the preparation of this work the author(s) used ChatGPT in order to check grammar and polish the wordings. After using this tool/service, the authors reviewed and edited the content as needed and take full responsibility for the content of the publication.

\section{Data availability statement}
The authors confirm that the data supporting the findings of this study are available within the article.


%
%
%




\bibliography{bibliography}

\end{document}